\documentclass[aps,pre,preprint,superscriptaddress,showpacs,showkeys]{revtex4-1}
\usepackage[english]{babel}
\usepackage[dviwin]{graphicx}
\usepackage[dvips]{color}
\usepackage{graphicx}% Include figure files
\usepackage{dcolumn}% Align table columns on decimal point
\usepackage{bm}% bold math
\usepackage{amsmath}
\usepackage{amscd}
\usepackage{amssymb}
\usepackage{tabularx}
\begin{document}
%\preprint{}
\title{Quantum Otto  cycle efficiency on coupled qudits}

\author{E. A. Ivanchenko}
\email{yevgeny@kipt.kharkov.ua,\,\,eaivanchenko1@gmail.com}
%\thanks{}

\affiliation{Institute for Theoretical Physics, National Science
Center \textquotedblleft{}Institute of Physics and
Technology\textquotedblright{},
 \\
  1, Akademicheskaya
str., 61108 Kharkov, Ukraine}
\date{\today}
\begin{abstract}
Properties of the coupled particles with spin 3/2 (quartits) in a constant magnetic field, as a working substance in the quantum Otto  cycle of the heat engine, are considered. It is shown that this system as a converter of heat energy  in work (i) shows the efficiency 1 at the negative absolute temperatures of heat baths, (ii) at the temperatures of the opposite sign the efficiency approaches to 1, (iii) at the positive temperatures of heat baths antiferromagnetic interaction raises efficiency threefold
in  comparison with uncoupled particles.
\end{abstract}
%05.70.-a Thermodynamics
%07.20.Pe Heat engines; heat pumps; heat pipes
%03.67.-a Quantum information
%03.65.Yz Decoherence; open systems; quantum statistical methods
%02.30.Yy Control theory
%
% \pacs {Visual imaging 87.63.L-, Visual perception,
% 42.66.Si, Nuclear magnetic resonance (NMR)
%in chemical physics, 82.56.-b, 03.67.-a Quantum information}
 % Visual imaging, 87.63.L-Visual imaging, 87.63.L-
%\keywords: Visualization, NMR, Quantum information
\maketitle
\section{Inroduction}\label{I}
As it is known,  thermodynamics has the broadest applications for  description of many physical phenomena \cite{TruesdellBharatha, TodaKuboSaito}. The quantum thermodynamics studies dynamics of heat and work in quantum systems. Researchers began to study quantum thermodynamic engines  after appearence of works \cite {ScovillSchulzDubois,GeusicSchulz-DuBoisScovill}. Thermodynamic cycles can be reformulated for quantum systems %\cite {Romero-Rochin}, %\cite {VilarRubi},
\cite {Kosloff1984,GevaKosloff,GevaKosloff1,tFeldmannrKosloff,Scully1, Scully2,RezekKosloff2006,QuanLiuSunNori,
ChotorlishviliToklikishviliBerakdar,HuangWangYi2013}. One of the important quantum thermodynamic cycles is an Otto  cycle.\\
Similarly to a classical Otto  cycle the quantum Otto  cycle  consists of two isochoric and two adiabatic stages too. A quantum isochoric process corresponds to a thermal exchange between a working body and thermal baths. During the quantum isochoric process only population of levels is reconstructed, whereas at the adiabatic process the working body produces work at the expense   of power level changes. The adiabatic process can be thermodynamic adiabatic or quantum adiabatic.
A process is thermodynamic adiabatic if the working substance is thermally isolated from a heat bath. However it does not exclude transitions  of purely quantum nature between levels while at the quantum adiabatic process   the population  density of levels is fixed.\\
The coupled spin systems
can  be used as quantum  thermodynamic engines. In   small systems with a finite number of degrees of freedom as finite-dimensional effects and quantum effects essentially influence thermodynamic properties of the system. The aim of this work is a research of quantum systems with a finite number of levels as a working substance in a quantum Otto  cycle, including viewing of the negative absolute temperatures \cite {Ramsey,HilbertHanggiDunkel,SwendsenWang}.
A  particle with spin-3/2 was studied both  in  the thermodynamic description in a stationary case in \cite{Ramsey} and in finite-time quantum thermodynamics in \cite{BirjukovJahnkeMahler}. The two  particles coupled by Heisenberg exchange interaction one of them with spin-3/2 and another with spin-l/2 in an  Otto  cycle are investigated in the  work \cite{AltintasMustecaplioglu2015}.\\
\indent The article is organised as  follows. In section II, the  properties of a working substance consisting of two coupled spins $3/2$ in a static magnetic field are described. In section III, a quantum heat Otto cycle is presented. Section IV   presents the results  graphically at the concrete control parametres. The final section V summarises the findings. In the Appendix auxiliary analytical formulae are presented.

\section {Working substance}\label {II}
The choice of a working substance for operation of the quantum heat engine is essential
% \cite {DunkelHilbert}
\cite {ThomasJohal,AzimiBerakdar,ZhangBarianiMeystre}.
The working substance in our case is featured by the Hamiltonian $\hat {H}$ of two coupled spins-$3/2$  (biquartit) with permutation symmetry of particles and the isotropic exchange interaction in a external static magnetic field $h$ applied along the z-axis:
\begin{equation}\label {eq:1}
\hat {H} =\mu h
   \left (E _ {4\times 4} \otimes S_3+S_3\otimes E _ {4\times 4} \right)
   +4J
   \overrightarrow {S} \otimes \overrightarrow {S},
   \end {equation}
in which $\mu$  is the quartit magnetic  moment.
The external control Hamiltonian  commutes with the internal interaction.
 $E _ {4\times 4} $ is the identity matrix, $S_1, \, S_2, \, S_3$ is the matrix representation  of components  of the spin 3/2 \eqref {eq:A1}, $J $ is the interaction constant. Cases $J > 0$ and $J <0 $ correspond to antiferromagnetic and ferromagnetic interactions,  respectively. \\
 %\cite {GeusicSchulz-DuBoisScovill}, \cite {FrenkelWarren}
    By the known partition function $Z $ \eqref {eq:A2} it is possible to calculate the free energy $F $, the entropy $S $, the internal energy $U $ and the heat capacity $C $:
\begin{equation}\label {eq:2}
  F =-1/\beta \ln Z, \, S =\beta^2 \partial_\beta F, \, U =\partial_\beta \beta F, \,
  C =-\beta^2 \partial _ {\beta \beta} \beta F,
  \end {equation}
 where $\beta=1/k_B T$ is the inverse temperature, $Z$ is the  partition function.
In the formulae \eqref {eq:2} the $ \beta $  partial differentiation idicates  that the other parametres are fixed.
  We use  units chosen so  that the magnetic moment $ $ equals 1, the Bolzmann constant $k_B$  equals 1, hence  $J, h, T$ in Joules.\\
According to  \cite {SchlienzMahler}, \cite {IJQI} the entaglement is determined by the values of  decomposition of the density  matrix   on the basis $ \rho \propto\sum _ {i=0} ^ {D-1} \sum _ {j=0} ^ {D-1} R _ {ij} C_i\bigotimes C_j $
\begin {equation} \label {eq:3}
 m _ {SM} = \sqrt {1 / (D-1) \sum _ {i=0} ^ {D-1} \sum _ {j=0} ^ {D-1} (R _ {i0} R _ {0j}-R _ {ij}) ^2},
\end {equation}
where $D $ is the qudit dimension of basis  $C_i $ (for a quartit $D=16$, for a qubit $D=4$) and $R _ {ij} $ are the components of the Bloch vector and $R _ {00}=1$.
\subsection {Local temperatures} \label {IIA}
For the Hamiltonian \eqref {eq:1}  the  quartit density matrix  $\varrho=\frac{1}{2\sqrt{5}}\sum _ {i=0} ^ {15}  R _ {i0} C_i$ and a local Hamiltonian is diagonal. The local entropy $s $ and the internal energy $u $ are defined by formulae
\begin {equation} \label {eq:4}
  s =-\sum _ {k=1} ^4\pi_k \ln \pi_k, \, u =\sum _ {k=1} ^4\varepsilon_k \pi_k,
\end {equation}
where the diagonal elements of the reduced density matrix $\varrho$ look like
\begin{equation*}
    \pi_1=1/20 (5-5
   p_1+p_2-5 p_3-5 p_4-5 p_5+5
   p_6+5 p_7+4 p_9-4 p _ {11} +15
   p _ {12}-5 p _ {13}-5
   p _ {14}-p _ {15} +5
   p _ {16}),
\end{equation*}
\begin{equation*}
    \pi_2=1/20 (
5 + p_1+3 p_2-5 p_3+5 p_4-5 p_5-5
   p_6+5 p_7-4 p_9+4 p _ {11}-5
   p _ {12}-5 p _ {13}-p _ {14} +7
   p _ {15} +5 p _ {16}),
\end{equation*}
\begin{equation*}
    \pi_3=1/20 (
5 + 3 p_1+p_2+5 p_3-5 p_4-5 p_5+5
   p_6-5 p_7-4 p_9+4 p _ {11}-5
   p _ {12} +5 p _ {13} +7
   p _ {14}-p _ {15}-5 p _ {16}),
\end{equation*}
\begin{equation*}
    \pi_4=1/20 (
5 + p_1-5 p_2+5 p_3+5 p_4+15 p_5-5
   p_6-5 p_7+4 p_9-4 p _ {11}-5
   p _ {12} +5 p _ {13}-p _ {14}-5
   p _ {15}-5 p _ {16}),
\end{equation*}
 and $\pi_ {i}$ are the populations of the reduced (local) diagonal quartit matrix  (it was provided with a choice of the  Hamiltonian \eqref{eq:1}). And only for  the diagonal matrix it is possible to determine the  local entropy correctly.
The eigenvalues $ \varepsilon_k $ of a local quartit Hamiltonian  are equal $h/2 (3,1,-1,-3) $. Formulae of level populations $p_i $ are given in the Appendix.
The local quartit temperature  is equal
\begin {equation} \label {eq:5}
   \beta _ {loc} = \frac {1} {T _ {loc}} = \frac {\partial s} {\partial u} = \frac {\partial s/\/\partial\beta} {\partial u/\/\partial\beta}.
\end {equation}
 The local temperature is not equal to the system temperature  of two coupled quartits $ \beta=1/T $ \cite {AzimiBerakdar}.
 We define the inverse spectroscopic temperature as \cite {GemmerMichelMahler}:
\begin {equation} \label {eq:6}
    \beta _ {Mloc} =-\left (1-\frac {\pi_1 +\pi_M} {2} \right) ^ {-1} \sum _ {i=2} ^ {M} \left (\frac {\pi_i +\pi _ {i-1}} {2} \right) \left (\frac {\ln\pi_i-\ln \pi _ {i-1}} {\varepsilon_i - \varepsilon_ {i-1}} \right),
\end {equation}
where $ \pi_i $  is the probability to find the quantum system at the energy $\varepsilon_i $, $M $  is the number of the highest
energy level $\varepsilon_M $, while the lowest one is labelled $\varepsilon_1$.
Actually it is a definition of the ensemble average of a random quantity $-\frac {\ln\pi_i-\ln \pi _ {i-1}} {\varepsilon_i - \varepsilon_ {i-1}} $ with the  distribution function density $ \left (1-\frac {\pi_1 +\pi_M} {2} \right) ^ {-1} \left (\frac {\pi_i +\pi _ {i-1}} {2} \right) $. \\
We  shall compare this expression for the local temperature  with the temperature definition \eqref {eq:5} in section IV.
\section {Heat Otto  cycle} \label {III}
We describe a working substance with a  Gibbs  quantum equilibrium distribution. The working substance passes through 4 stages of the Otto cycle. Therefore in  the theoretical description of this Gedankenexperiment the temperature of baths and  the temperature of the working substance are equal because of the assumption about quasi-stationarity  and adiabaticity. The power as a  done  work for infinite time   is equal to zero. In the time-finite description at nonadiabaticity the efficiency is less, but the  power produced will be distinct from zero. \\If to consider  the equation for a density matrix of  the  working substance
   $\partial_t\rho=-i[\hat {H},\rho]+L \rho$,
where a  Lindblad   operator $L$ takes into account the environmental  influence, then in a stationary case $\partial_t\rho=0$, and the environmental influence means  that for a working substance temperature $T$ (in the first stage) or $T'$  (in the third stage)  is established. In the assumption of a weak environmental influence of the  term $L\rho$ is  little and  we obtain  the equation $[\hat {H},\rho]=0$ for a density matrix $\rho$.  The solution for the density matrix is any function depending on the Hamiltonian $\hat {H}/T$, where  $T$  is the temperature. The Gibbs quantum distribution is  followed  from the requirement $\rm Tr \rho =1$.\\
Let's feature 4 stages of a quantum quasi-static Otto cycle \cite {ThomasJohal}. \\
\emph {Stage} 1: the system of two coupled quartits in a magnetic field $h $ attains  thermodynamic equilibrium with  a heat bath of temperature $T $. The occupation probabilities are  determined by temperature $T $ and a magnetic field $h $. Thus the occupations  change, and the energy levels do not change. The work is not produced during this  isochoric process, and the working substance absorbs heat $ Q_1$ from the  bath:
\begin {equation} \label {eq:7}
 Q_1 =\sum _ {i=1} ^ {16} e_i (p_i-p '_i).
\end {equation}
\emph {Stage} 2: the system is isolated from the heat bath and the magnetic field  is changed from $h $ to $h ' $  by an adiabatic process, and the energy levels slowly change. Accoding to the adiabatic theorem the occupation probabilities  of each energy level maintain. The work is produced:
\begin {equation} \label {eq:8}
  W_2 =\sum _ {i=1} ^ {16} p_i (e ' _i-e_i).
\end {equation}
\emph {Stage} 3: the system is brought in  contact with a heat bath at temperature $T ' $. Upon attaining thermodynamic equilibrium with the bath the occupation probabilities are determined by temperature $T ' <T $ and a magnetic field $h ' $. The system gives off heat  energy $Q_3$ to the bath:
\begin {equation} \label {eq:9}
 Q_3 =\sum _ {i=1} ^ {16} e ' _i (p ' _i-p_i).
\end {equation}
\emph {Stage} 4: the system is removed from a cold bath and undergoes another adiabatic process which changes the magnetic field from $h ' $ to $h $ but keeps the occupation probabilities unaffected. The energy levels slowly change and the work $W_4$ is produced:
\begin {equation} \label {eq:10}
  W_4 =\sum _ {i=1} ^ {16} p '_i(e_i- e '_i).
\end {equation}
The system is brought in contact with a heat bath at temperature $T $. Heat is absorbed from the bath and the system returns to its initial state.\\
Note that Q > 0 means that heat is absorbed from the bath by the system, and W > 0 means that work is done on the system and the opposite for the opposite direction of the inequalities.\\
At non-adiabatic transitions fast dynamics on adiabatic branches is responsible for frictional losses. As a result the system is incapable to follow adiabatically along time-dependent changes of the Hamiltonian. The deviation from the quantum adiabatic behavior is expressed by losses which appear from generation of inertial components on adiabatic curves and their dephasing on isochores  \cite {FeldmannKosloff2009}.
All the cycle is presented on the diagramme \eqref {eq:11}.
\begin{equation}\label{eq:11}
\begin{CD}
    1                      @>Q_1>>             2           \\
    @AW_4AA                                     @VVW_2V    \\
    4                      @<<Q_3<              3
\end{CD}
\end{equation}
The energy change during the cycle is equal to zero:
\begin {equation} \label {eq:12}
    Q_1+W_2+Q_3+W_4=0.
\end {equation}
The heat transfered in \emph {Stage} 1 and in \emph {Stage} 3 respectivly is
\begin{equation} \label {eq:13}
  Q_1 =\sum _ {i=1} ^ {16} e_i (p_i-p ' _i) =J m+h n, \,
  Q_3 =\sum _ {i=1} ^ {16} e ' _i (p ' _i-p_i) =-J m-h'n,
\end{equation}
where
\begin{subequations}\label {eq:14}
\begin{eqnarray}
m= -11 \left(p_1-p_1'+ p_2-p_2'+ p_9-p_9'\right)+\nonumber\\
9 \left(
   p_5-p_5'+
   p_{11}-p_{11}'+
   p_{12}-p_{12}'+
   p_{13}-p_{13}'+
   p_{14}-p_{14}'+
   p_{15}-p_{15}'+
   p_{16}-p_{16}'\right)-\nonumber\\3 \left(
   p_3-p_3'+ p_4-p_4'+
   p_6-p_6'+ p_7-p_7'+
   p_{10}-p_{10}'\right)-15
   \left(p_8-p_8'\right),\\
    n= -p_1+p_1'+p_2-p_2'-p_4+p_4'+p_6-p_6'-p_{14}+p_{14}'+p_{15}-p_{15}'+\nonumber\\
2(-p_3+p_3'+p_7-p_7'-p_{13}+p_{13}'+p_{16}-p_{16}')+3(-p_5+p_5'+p_{12}-p_{12}').
\end{eqnarray}
\end{subequations}
The work is done in Stage 2 and Stage 4 when the energy levels change at the fixed occupation probabilities.
Due to energy level changes the work done by the quantum heat engine is
\begin {equation} \label {eq:15}
W _ {out} = W_2 + W_4 = (h'-h) n,
\end {equation}
where
\begin {equation} \label {eq:16}
W_2 =\left (h-h' \right) \left (p_1-p_2+2 p_3+p_4+3
   p_5-p_6-2 p_7-3 p _ {12} +2
   p _ {13} +p _ {14}-p _ {15}-2
   p _ {16} \right),
\end {equation}
\begin {equation} \label {eq:17}
W_4 =\left (h '-h\right)
   \left (p_1 '-p_2 ' +2 p_3 ' +p_4 ' +3
   p_5 '-p_6 '-2 p_7 '-3 p _ {12} ' +2
   p _ {13} ' +p _ {14} '-p _ {15} '-2
   p _ {16} '\right).
\end {equation}
 The efficiency of transformation of heat into work at $Q_1> 0, \, Q_3 <0$ is
\begin {equation} \label {eq:18}
\eta =\frac {W _ {out}} {Q _ {in}} = \frac {- (W_2+W_4)} {Q_1} = \frac {\eta_0} {1+\frac {J m} {h n}}.
\end {equation}
It is obvious that the interaction between particles can give both the enhancement and the reduction of the efficiency concerning noninteracting particles.
For uncoupled particles that is $J=0$ the efficiency is $ \eta_0=1-\frac {h '} {h} $ \cite {Kieu}.
\subsection {Local description} \label {IIIA}
In this subsection, following the article \cite {ThomasJohal},
 we feature how the individual quartits undergo the cycle. Heat, transfered locally between one quartit and a heat bath, is
\begin {equation} \label {eq:19}
   q_1 =\frac {h} {2} n, \, q_2 =-\frac {h '} {2} n,
\end {equation}
for hot and cold baths accordingly. The work done by one particle is
\begin {equation} \label {eq:20}
    w=q_1+q_2 =\frac {h-h '} {2} n.
\end {equation}
\begin {equation} \label {eq:21}
  W=2w = (h-h ') n.
\end {equation}
Thus the total performed work is the sum of local work obtained from each qudit. It is a consequence of permutation symmetry of the hamiltonian. \\
The total heat, absorbed (produced) by the system in \emph {Stage} 1 (\emph {Stage}  3) can be written as
\begin {equation} \label {eq:22}
  Q_1=J m+2 q_1, \,
  Q_2 =-J m-2 q_2.
\end {equation}
  \section {Results} \label {IV}
\begin {figure} [tbp]
\includegraphics [width = 3 in] {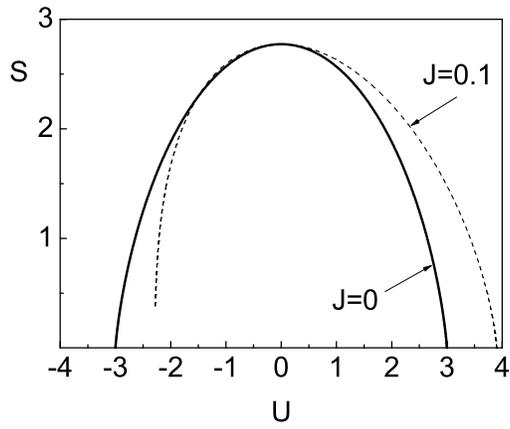}
\caption {\label {SU_parametric_beta}  Entropy $S $ is plotted as a function of  the internal energy $U $ for  a biquartit  at the fixed magnetic field $ h=1$.}
\end {figure}

\begin {figure} [tbp]
\includegraphics [width = 3 in] {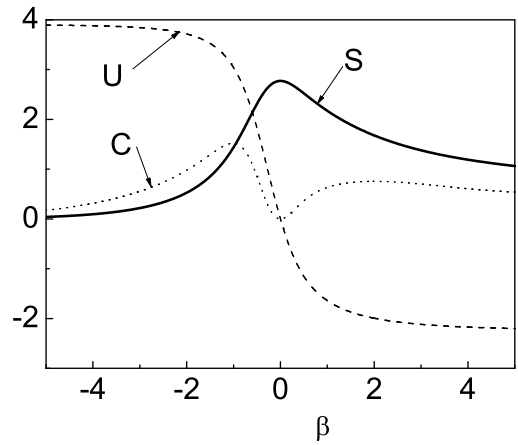}
\caption {\label {USC}The internal energy $U $, the entropy $S $ and the heat capacity $C $ are plotted as  functions on the inverse temperature $\beta$ for $ h=1, \, J=0.1$. At $ J=0$ the entropy and the heat capacity are the even functions, the internal energy is the odd function \cite {Ramsey}.}
\end {figure}

\begin {figure} [tbp]
\includegraphics [width = 3 in] {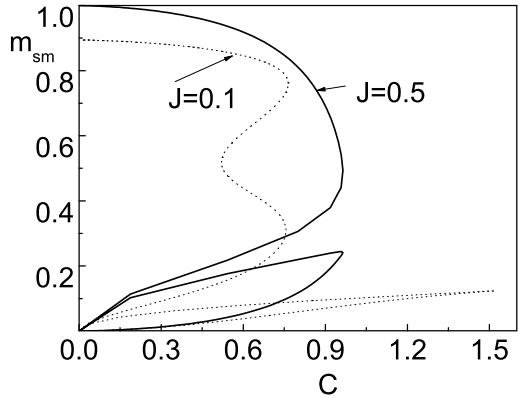}
\caption {\label {Entangl} Parametric dependence of the  entanglement $m_{SM} $ and the heat capacity $C $  on the inverse temperature at $h=1$ for different coupling constants. The closed parts of graphs correspond to the negative temperature, and  unclosed  ones to the positive temperature.}
\end {figure}

\begin {figure} [tbp]
\includegraphics [width = 3 in] {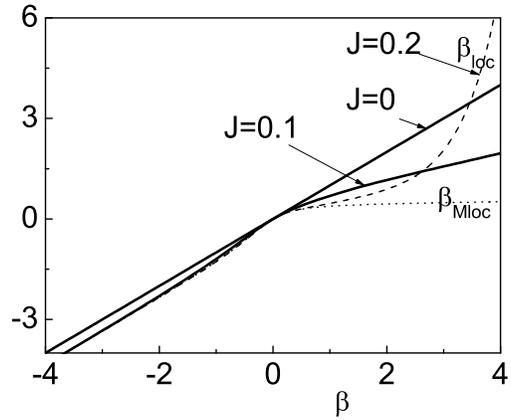}
\caption {\label {tloc} Inverse local temperature of the quartit $ \beta _ {loc} $ versus the inverse temperature of the biquartit $ \beta $ for
$h=2, \, J=0, \, 0.1, \, 0.2$.}
\end {figure}

\begin {figure} [tbp]
\includegraphics [width = 3 in] {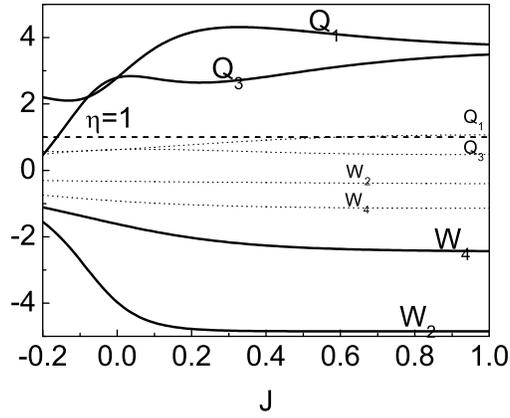}
\caption {\label {mm_effic_1} Dependence of heat and work on a coupling constant at all stages of the Otto cycle at $T =-1, \, T ' =-3, \, h=1, \, h ' =-1$. The dashed line is   the  efficiency of the heat energy conversion in work. Hereinafter the bold lines are for the biquartit; the pointwise lines are for the biqubit.}
\end {figure}

\begin {figure} [tbp]
\includegraphics [width = 3 in] {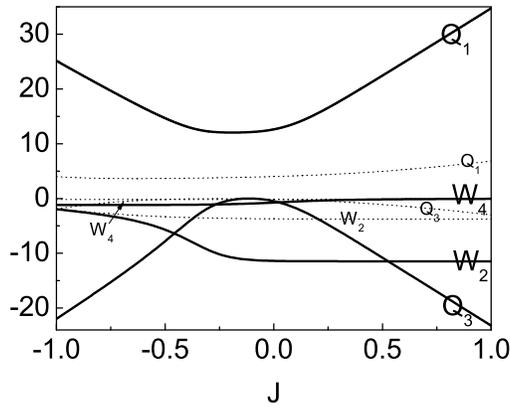}
\caption {\label {mpWwQq} Heat and work versus the coupling  constant for  $ T =-1, \, T ' =2, \, h=4, \, h ' =0.155$.}
\end {figure}

  We illustrate analytical results graphically at the control parameters $(h, h', J, T, T' )$.  These parameters characterise the working substance  at the different stages. \\
  \indent \textit {Working substance}.
  Fig.~\ref {SU_parametric_beta} shows the entropy dependence on the internal energy. Dependences of the internal energy, the entropy and the heat capacity on inverse temperature are shown in Fig.~\ref {USC}. Coupling of spins breaks the symmetry which is at the equidistant disposition of energy levels in the system \cite {Ramsey}. Dependence of the entanglement on the heat capacity is presented in Fig.~\ref {Entangl}. The dependence singularity is based on the fact that at a small constant $J $ is multivalued, and at a big one it is two-valued. Numerical comparison of dependence of local inverse temperature definitions from the inverse
system temperature is given in Fig.~\ref {tloc}. Both definitions give the same results at a small interaction constant. At $J=0, \, 0.1, \, 0.15$ there is the full coincidence of definitions of local temperatures both at negative and positive $ \beta $ (bold lines). At $J> 0.17$ there is an appreciable discrepancy at positive $ \beta $. \\
At $ J <0$ the divergence is observed at negative temperatures, that is the graphs are symmetric concerning the origin of coordinates. These divergences both at $J <0$ and at $J> 0$ are caused by    the energy level perturbation.\\
       We describe the efficiency of a quantum Otto cycle on coupled quartits for some possible sets of positive and negative signs of the quantities $Q_1, W_2, Q_3, W_4$. Reduced letters in plots show the characteristics of two coupled qubits (biqubit) for comparison with the paper \cite{ThomasJohal}. It is possible due to control parametres $h, \, h ', \, J. $ It is obvious that at $h=h ', \, J \neq 0, \, Q_1> 0, \, W_2=0, \, Q_3 <0, \, W_4=0$, that is heat is just transfered from a hot bath into a cold one. \\
   \indent \textit {The quantum heat engine between baths with negative absolute temperatures}.
   At negative temperatures of heat baths $T <0, \, T ' <0, \lvert T\rvert <\lvert T '\rvert $ the situation, when $Q_1> 0, \, Q_3> 0, \, W_2 <0, \, W_4 <0$ (see Fig.~\ref {mm_effic_1}) is possible. In this case, the efficiency of convertion of heat in work is equal to 1 \cite {Ramsey}, according to \eqref{eq:12} and the efficiency definition $ \eta =\frac {W _ {out}} {Q _ {in}} = \frac {- (W_2+W_4)} {Q_1+Q_3} =1. $ \\
     \indent \textit {The quantum heat engine between baths with absolute temperatures of the opposite sign.} \\
       At temperatures of baths $T <0, \, T '> 0$ the situation, when $Q_1> 0, \, Q_3 <0, \, W_2 <0, \, W_4 <0$ (see Fig.~\ref {mpWwQq} ) is possible. In this case the efficiency of conversion of heat in work is equal $ \eta =\frac {- (W_2+W_4)} {Q_1} $ \eqref {eq:18} and because of a small leakage it approaches unity, as shown in Fig.~\ref {Eff_mp}. A shift of the maximum efficiency in the biquartit in regard to the biqubit is observed. \\
    At a modification of driving parameters the efficiency can exceed more than three times the efficiency of uncoupled 3/2-spins, as seen from Fig.~\ref {Eff082qutrit2qubitpp}. For the biqubit the maximum efficiency is moved towards the increase of the interaction constant \cite {ThomasJohal}. \\
 \indent \textit {The quantum heat engine of conversion of work in heat between baths with the positive temperatures}.
 At the positive temperatures of baths $T> 0, \, T '> 0, \, T> T ' $ the situation, when $Q_1 <0, \, Q_3 <0, \, W_2> 0, \, W_4> 0$ (see Fig.~\ref {pp_effic_1w_to_heat}) is possible. In this case the efficiency of conversion of work in heat equals $ \eta =\frac {- (Q_1+Q_3)} {(W_2+W_4)} =1. $
  At some set of parametres $Q_1$ changes its sign, then the total work $W_2+W_4$ changes its sign, that is in a neighbourhood $Q_1=0$ the work done over the system, is entirely converted in heat \cite{BirjukovJahnkeMahler}. \\
\indent \textit {The work as a function of  the entanglement in  a  biquartit}.
The entanglement $m_{SM}$ and the work  $W_2, W_4$ in the Otto cycle   are determined with the matrices  $\rho(e_i/T)$, $\rho'(e_i'/T')$  respectively.   Fig.~\ref{W2W4}  shows the  parametric dependence of the work as a function of the entanglement. It is evident that the work increases  along with the increase of entanglement at $J<0$  in the second  and  fourth  stages,  at  $J>0$ the work decreases with the increase of entanglement. In the absence of interaction the entanglement is equal to zero.\\
\indent \textit {The total work per cycle $-(W_2+W_4)$ as a function of  the magnetic field $h'$ in  a  biquartit}.
In the considered approach in limit of small systems with only a few degrees of freedom the necessary condition for a heat engine $h '/T '> h/T $, for a  refrigerator $h '/T ' <h/T $. In a multilevel system as it was marked in \cite {UzdinKosloff2014}, it is difficult to find simple criteria to answer when the Otto cycle is a heat engine, and when it is a refrigerator. At some parameters these criteria (see Fig. 11) are carried out, and at others are not.\\
For $J=0$ below the Carnot point (left vertical line, $h'/h = T'/T $) the device acts as a refrigerator and above it until $h' =h $ it performs as an engine. For $h'> h $ the device performs as a heater as it takes work to make the cold bath hotter.
 At $J <0$ the done  work decreases and the Carnot point slightly moves to the left. At positive $J> 0$ the done  work decreases and the Carnot point moves to the right before coincidence with  the  point $h'=h $ at $J=0.2$. In this case the device works as a refrigerator ($h'<h $) or as a heater ($h'> h $).
 \section {Conclusion} \label {V}
\begin {figure} [tbp]
\includegraphics [width = 3 in] {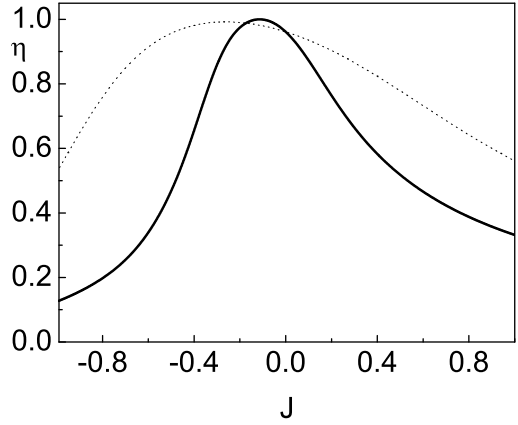}
\caption {\label {Eff_mp} Efficiency of the conversion of heat in work with parameters as in Fig. ~\ref {mpWwQq}. The heat leakage in the biqubit equals \,\,-0.0028; the leakage in the biquartit is $Q_3 =-0.0021$. The maximum efficiency equals 0.999 in  the biqubit for $J = - 0.26$, and in the biquartit for $J = - 0.11$.}
\end {figure}

\begin {figure} [tbp]
\includegraphics [width = 3 in] {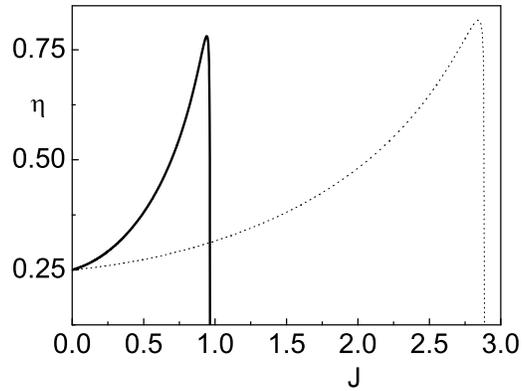}
\caption {\label {Eff082qutrit2qubitpp} Efficiencies $ \eta $ of the biqubit and the biquartit transformations of heat in work depending on the coupling constant $J $ at $T=2.5, \, T ' =0.25, \, h=16, \, h ' =12. $ The Carnot limit is 0.9.}
\end {figure}

\begin {figure} [tbp]
\includegraphics [width = 3 in] {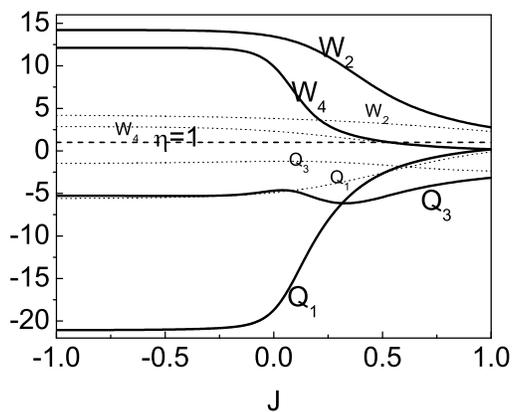}
\caption {\label {pp_effic_1w_to_heat} Heat and work versus the coupling constant for $ T=2, \, T ' =1, \, h=4, \, h ' =-1$. The dashed line is the efficiency of conversion of work in the heat energy.}
\end {figure}

\begin {figure} [tbp]
\includegraphics [width = 3 in] {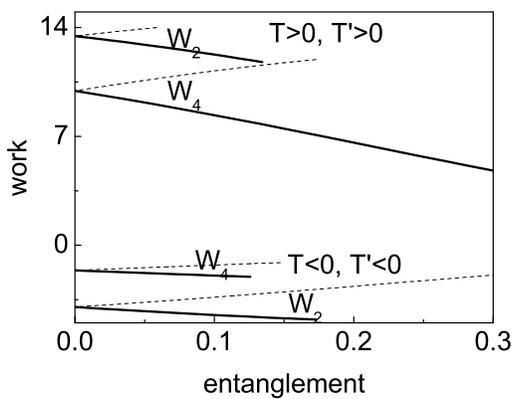}
\caption {\label {W2W4} Work $W_2<0, W_4<0 $ versus the entanglement $m_{SM}$  with parameters as in
Fig.~\ref {mm_effic_1}; at $W_2>0, W_4>0$   parameters as in Fig.~\ref{pp_effic_1w_to_heat}. Full lines correspond to $J>0 $, dashed ones $J<0 $.}
\end {figure}

The quasi-stationary quantum Otto cycle, when the working body is the coupled system of two 3/2-spins, being in a magnetic field, is explored. Some performances of the quantum Otto cycle on the coupled spins, generated by various sets of driving parameters, are considered. The analysis of possible quantities of the cycle efficiency depending on driving parameters is carried out. There are the restrictions on driving parameters $T> T ', \, \, h '/T '> h/T $ for the conversion of heat in work (see Fig.~\ref {mm_effic_1},  \ref {mpWwQq},  \ref {Eff_mp},  \ref {Eff082qutrit2qubitpp}).
It is shown that the efficiency of conversion of heat in work at negative temperatures of heat baths equals 1, at temperatures of the opposite sign it approaches 1. At positive temperatures of heat baths the antiferromagnetic interaction of spins \cite{ThomasJohal} raises the efficiency more than threefold in comparison with uncoupled spins \cite {ThomasJohal}. The dependence on a system size is revealed in a displacement of the maximum efficiency in regard to the enhancement of the interaction constant (see Fig.~\ref {Eff_mp}, \ref {Eff082qutrit2qubitpp}).\\ Dependence of  the work on entanglement with the limiting values of efficiency in cases of  conversion of  heat into work  and  work into heat  is presented.\\
When dealing with realistic systems, Quantum Thermodynamics introduces finite time  in the analysis. For the Carnot, Otto, Stirling and other cycles time is introduced on all stages.
The pioneering  studies in finite time  quantum thermodynamics in the  method of quantum generators of open systems were done  by R. Kosloff and co-workers in works \cite {Kosloff1984, GevaKosloff,TovaFeldmannRonnieKosloff,PalaoKosloffGordon2001,FeldmannKosloff2012, Kosloff2013}.
%On the %isochores it can be a simple master equation with constant %transition rates from level to level. On the adiabats it can be zero %time (which will leave the level population unchanged, for no time to %redistribute the population), or a choice of a function $h (t) $ which %would cause population change.
Nowadays other approaches \cite {HenrichMahlerMichel, BirjukovJahnkeMahler,IvanchenkoLviv, Popescu2014} are    actively developed. It is necessary also to define  more exactly   the quantum thermodynamical  work and heat in order to study local effective dynamics in microsystems \cite {AlickiHorodecki2004,SchroderMahler2008epl}.
 \subsection*{Acknowledgements}
The author is thankful to anonymous referees for many helpful remarks.
\appendix
\section{}
The matrix representation of a vector of a spin 3/2 looks like
\begin{equation}\label{eq:A1}
S_1=\left(
\begin{array}{cccc}
 0 & \frac{\sqrt{3}}{2} & 0 & 0 \\
 \frac{\sqrt{3}}{2} & 0 & 1 & 0 \\
 0 & 1 & 0 & \frac{\sqrt{3}}{2} \\
 0 & 0 & \frac{\sqrt{3}}{2} & 0 \\
\end{array}
\right),\,S_2=\left(
\begin{array}{cccc}
 0 & -\frac{i \sqrt{3}}{2} & 0 & 0
   \\
 \frac{i \sqrt{3}}{2} & 0 & -i & 0
   \\
 0 & i & 0 & -\frac{i \sqrt{3}}{2}
   \\
 0 & 0 & \frac{i \sqrt{3}}{2} & 0
   \\
\end{array}
\right),\,S_3=\left(
\begin{array}{cccc}
 \frac{3}{2} & 0 & 0 & 0 \\
 0 & \frac{1}{2} & 0 & 0 \\
 0 & 0 & -\frac{1}{2} & 0 \\
 0 & 0 & 0 & -\frac{3}{2} \\
\end{array}
\right).
\end{equation}
The density matrix on the basis of eigenfunctions of the Hamiltonian (the Gibbs representation) looks like
\begin {equation} \label {eq:A2}
  \rho=e ^ {-\beta \hat {H}}/Z =\sum _ {i=1} ^ {16} p_i P_i,
\end {equation}
where $ \beta=1/T $ is the inverse temperature, \, $Z = {\rm
Tr \,} e ^ {-\beta \hat {H}} = \sum _ {i=1} ^ {16} e ^ {-\beta e_i} $ is the partition function, $p_i=e ^ {-\beta e_i}/Z $ are the occupation densities, $P_i = | e_i> <e_i |, \, P_i P_k =\delta _ {i, k} P_i $ are the projectors constructed of eigenvectors of the Hamiltonian $ |e_i> $, corresponding to the eigenvalues $e_i (h, J) \equiv e_i $; \,
$ e_1 =-h-11J, \, e_2=h-11J, \, e_3 =-2h-3J, \, e_4 =-h-3J, \, e_5 =-3h +9J, \, e_6=h-3J, \, e_7=2h-3J, \, e_8 =-15J, \, e_9 =-11J, \, e _ {10} =-3J, \, e _ {11} =9J, \, e _ {12} =3h +9J, \, e _ {13} =-2h+9J, \, e _ {14} =-h+9J, \, e _ {15} =h+9J, \, e _ {16} =2h+9J, \, \sum _ {i=1} ^ {16} e_i=0, \, e ' _i (h ', J) \equiv e ' _i $. The normalized eigenvectors equal:
\begin {subequations}
\begin {eqnarray}
|e_1> = \sqrt {10/3} (0_7,1,0_2,-\frac {2} {\sqrt {3}}, 0_2,1,0_2), \,
|e_2> = \sqrt {10/3} (0_2,1,0_2,-\frac {2} {\sqrt {3}}, 0_2,1,0_7), \nonumber\\
|e_3> =1/\sqrt {2} (0 _ {11},-1,0_2,1,0), \,
|e_4> =1/\sqrt {2} (0_7,-1,0_5,1,0_2), \,
|e_5> = (0 _ {15}, 1), \nonumber\\
|e_6> =1/\sqrt {2} (0_2,-1,0_5,1,0_7), \,\nonumber
|e_7> =1/\sqrt {2} (0,-1,0_2,1,0 _ {11}), \\\nonumber
|e_8> =1/2 (0_3,-1,0_2,1,0_2,-1,0_2,1,0_3), \,
|e_9> = \sqrt {20}/3 (0_3,1,0_2,-\frac {1} {3}, 0_2,-\frac {1} {3}, 0_2,1,0_3), \\\nonumber
|e _ {10}> =1/2 (0_3,-1,0_2,-1,0_2,1,0_2,1,0_3), \,
|e _ {11}> =1/\sqrt {20} (0_3,1,0_2,3,0_2,3,0_2,1,0_3), \\\nonumber
|e _ {12}> = (1,0 _ {15}), \,
|e _ {13}> =1/\sqrt {2} (0 _ {11}, 1,0_2,1,0), \,
|e _ {14}> =1/\sqrt {5} (0_7,1,0_2, \sqrt {3}, 0_2,1,0_2), \\\nonumber
|e _ {15}> =1/\sqrt {5} (0_2,1,0_2, \sqrt {3}, 0_2,1,0_7), \,
|e _ {16}> =1/\sqrt {2} (0,1,0_2,1,0 _ {11}),
\end {eqnarray}
\end {subequations}
where $0_k\equiv\overbrace {0,0,..., 0} ^ {k \, times} $. \\
The biqubit Hamiltonian has the same structure as in the equation \eqref {eq:1} with the eigenvalues $-3 J, J,-h + J, h + J $.
\begin {figure} [tbp]
\includegraphics [width = 3 in] {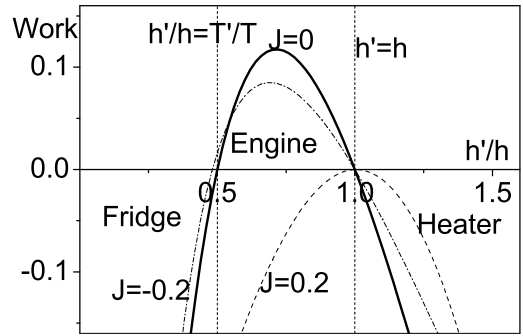}
\caption {\label {AxisW24} Work   versus the control parameter  $h'$    at $h=1, T=1, T'=0.5$. Full line correspond to $J=0; $ dashed,
dashed-dot ones correspond $J=0.2, J=-0.2$ respectively.}
\end {figure}
\clearpage
% \bibliographystyle {alpha}
% \bibliographystyle {unsrt}
%\bibliographystyle {apsrev}
% \bibliographystyle {gost780u}
%\bibliography {Negativ_Temp}
%%\bibliography {PhysEB11387}
%%\end{document}
%
\end{document}